\documentstyle[11pt]{article}
\date{}
\def\ifmath#1{\relax\ifmmode #1\else $#1$\fi}%

\def\ln{\ifmath{{\mathrm{ln}}}}
\def\eff{\ifmath{{\mathrm{eff}}}}
\def\cal{\ifmath{{\mathrm{cal}}}}

\begin{document}
\null\vskip-2cm
\hskip310pt HZPP-96-06

\hskip310pt June 15, 1996
\bigskip

\begin{center}
{\Large A Possible Signal for QGP Formation} 
\end{center}
\begin{center}
{\Large from the Minimum-Bias Data of Relativistic}
\end{center}
\begin{center}
{\Large Heavy Ion Collisions }
\end{center}
\bigskip
\begin{center}
{Liu Lianshou\ \ \ \ Hu Yuan \ \ \ \ Deng Yue\\
{\small Institute of Particle Physics, Huazhong Normal University,
         Wuhan 430070, China}}
\end{center}
\vskip1cm

\begin{center}
\begin{minipage}{125mm}
\begin{center}{\bf Abstract }\end{center}
{It is argued that the experimentally observed strong upward-bending of
the logarithm of factorial moments versus that of phase space
partition number in the 
higher-dimensional phase space of nucleus-nucleus collisions is due to the 
superposition of elementary collision processes in these collisions.
A direct implication of this observation is that, at high enough energy and/or 
density, when the produced particles from individual elementary processes 
are melted into a unique system the above-mentioned superposition effect
will disappear and the factorial moments will not be strongly upward-bending
any more. So, the disappearance of strong upward-bending of 
higher-dimensional factorial moments
in heavy ion collision may be taken as a signal for 
the formation of a unique system, or QGP, in this collision.}

\vskip0.5in

Signal of QGP formation \ \ Scaling behaviour of factorial moments \ \ 

\end{minipage}
\end{center}

\vfill

\begin{center}
{\bf INSTITUTE OF PARTICLE PHYSICS\\
HUAZHONG NORMAL UNIVERSITY\\
WUHAN \ \ \ CHINA}
\end{center}

\newpage

New powerful high energy heavy ion colliders are in construction both in
the United States and in Europe with the goal of formating a new state of
matter --- the quark gluon plasma QGP. Various signals for the formation of 
QGP have been proposed\cite{s1}, but in view of the complexity of the problem, 
further investigation in this respect is worthwhile. To propose a new signal
for the formation of QGP using the minimum-bias data of relativistic heavy
ion collisions is the goal of this paper.

As is well known, heavy ion collision at conventional energies can be regarded
as the superpostion of a large number of elementary collision processes. It
is predicted that at high enough energy and/or density the particles produced
in the individual elementary processes will be thermalized and melted 
into a unique system, formating a new state of matter --- QGP. 
So, if we are able to find a characteristic phenomenon that is connected
with the superpostition of individual collision processes,
then the disappearance of this phenomenon will indicate that 
a unified system, or QGP, has been formed. In the following a
new signal for the formation of QGP is proposed along this line of thought.

The phenomenon to be used concerns the scaling behaviour of 
higher-dimensional factorial moments (FM).  
The $q$th order factorial moment $F_q$ is defined as\cite{s2}
\begin{equation}
\label{e1}
F_q(M)={\frac {1}{M}}\sum\limits_{m=1}^{M}{{\langle n_m(n_m-1)\cdots
(n_m-q+1)\rangle }\over {{\langle n_m \rangle}^q}},
\end{equation}
where a region $\Delta$ of phase space is divided into 
$M$ cells, $n_m$  is the multiplicity in the $m$th cell, and
$\langle\cdots\rangle$ denotes vertically averaging over many events.

The scaling behaviour of FM is observed in experiments by ploting the 
ln$F_q$ versus ln$M$. It turns out that                           
in 1-D phase space, especially in (pseudo)rapidity
variable, ln$F_q$ for nucleus-nucleus collisions are in general rising more
slowly as the increasing of ln$M$ than that in hadron-hadron collisions, 
and the heavier the colliding nuclei are, the weaker is 
the rising of ln$F_q$\cite{s3}. This nuclear effect can be explained 
as due to the superposition of elementary
nucleon-nucleon collisions in a nucleus-nucleus interaction\cite{s4}.

However, it is interesting that in the 2-D($\eta$,$\phi$) and 3-D
($\eta$,$p_t$,$\phi$) space, the log-log plots of FM versus $M$ for
nucleus-nucleus collisions turn 
out to be strongly bending upwards\cite{s5,s6}, 
much more
stronger than what is observed in $e^+$-$e^-$ and hadron-hadron collisions
\cite{s7}-\cite{s10}.  

A natural question is: What is the origin of the strong upward bending of FM in
higher-dimensional phase space of nucleus-nucleus collisions?
Is it also a manifestation of nuclear effect, i.e. is it also
due to the superposition of the contributions from elementary processes?

In order to answer this question let
us recall that in calculating the higher-dimensional FM 
how to divide the phase space in different directions is essential\cite{s11}.
A quantity $H$ called Hurst exponent can be introduced to characterize the
way of phase space partition. It is defined as
\begin{equation} \label{e2}
H_{ab}= {{\ln}M_a\over {\ln}M_b},
\end{equation}
where $M_a$ and $M_b$ are the partition numbers in the directions $a$
and $b$ respectively. The anomalous scaling of FM, if exists, is 
definitely connected with a certain value of $H$. In analyzing 
experimental data, only when the FM's are calculated with the 
correct value of $H$, i.e. using the correct way of phase space partition,
a straight line in ln$F_q$ vs. ln$M$ can be observed.
On the other hand, if the way of phase space partition is incorrect, i.e.
if the FM's are not calculated with the right value of $H$, the resulting
ln$F_q$ vs. ln$M$ will be bending upwards\cite{s11}.

Conventionally, in calculating FM one usually takes the same partition number 
in different phase space directions, i.e. when the phase space
is divided into $M_a$ pieces in direction $a$, it is divided into $M_b=M_a$
pieces in direction $b$. This is called ``self-similar'' analysis, cf. Fig.1a.
As pointed out in Ref.[11], due to the highly anisotropy of phase space 
the right way of analysis should be to take different partition number in 
different direction: $M_b\not= M_a$. This is called self-affine analysis, 
cf. Fig.1b.  The correct value of $H_{ab}$ for this self-affine analysis can be 
obtained by fitting the 1-D plots of ln$F_2$ vs. ln$M$ to the Ochs saturation 
formula\cite{s12}.

The above assertion has been tested in hadron-hadron collision
experiment\cite{s13}.  The Hurst exponent $H_{ab}$ is found
to be less than unity when $a$ is the longitudinal direction ($y$ or $\eta$)
and $b$ is the transverse direction($p_t$ or $\phi$). 
This explains the slightly upward-bending of 3D FM in the conventional 
self-similar analysis of hadron-hadron collision data.

However, \ the strong upward-bending of \ higher-dimensional FM \ observed
in nucleus-nucleus collison experiments cannot be explained
with only this reasoning. The problem is: How can the
superposition of contributions from individual elementary
processes make the higher-dimensional FM 
in nucleus-nucleus collsions bending upwards so strongly, while the same
superposition weakens the rising of 1-D FM in these collisions. 

This striking result can be understood as the following. In a
nucleus-nucleus collision process due to the fluctuation of colliding
parameters the rapidity centers of individual elementary collisions
do not coincide but are scattered randomly on the rapidity axis. When the
(pseudo)rapidity region of each elementary collision is divided 
into $M_\parallel$
pieces, their superposition makes the whole rapidity region be divided into a
much larger number ($M_\parallel^{\eff}$) of pieces, cf. Fig.2.
On the other hand, similar effect does not exist in the transverse direction 
$\perp$ ($p_t$, $\phi$) where the phase space region is the same for
all the elementary collisions and their superposition makes $M_\perp$ 
no change. Therefore, although the actual Hurst exponent $H_{\parallel\perp}$ 
for each elementary collision is less than unity the effective Hurst 
exponent for their superposition will be much greater than this value,
\begin{equation}
\label{e3}
H_{\parallel\perp}^{\eff}= {M_\parallel^{\eff}\over M_\perp}\gg 
{M_\parallel\over M_\perp}=H_{\parallel\perp}.
\end{equation}
So, when we calculate FM self-similarly, i.e. with 
$H_{\parallel\perp}^{\cal}=1$,
the resulting ln$F_q$ vs. ln$M$ will be bending strongly upwards.

In order to demonstrate the above arguement we have made a simple model. In
this model each elementary collision in the ``nucleus-nucleus collision'' 
with $A$ nucleons is simulated by a two-dimensional random cascading $\alpha$ 
model\cite{s11}, in which the regions in direction $b$ for all the $A$ 
elementary collisions are the same, while that in direction $a$ for the $i$th 
elementary collision is placed randomly at [$-\omega \cdot r_i$, 
$1+\omega - \omega\cdot r_i$], with $r_i$ a random number 
in [0,1] and $\omega> 0$ a fixed parameter. The resulting
particles from all the $A$ elementary $\alpha$ models are then superposed 
together to form a ``nucleus-nucleus collision event''. 

The results of ln$F_2$ vs. ln$M$ for $A=2$--5 are shown in Fig.3. They are all 
bending upwards, and as the number $A$ of ``elementary collision'' increases
the upward-bending of ln$F_2$ vs. ln$M$ becomes stronger and stronger.
Thus the model Monte-Carlo simulation confirms the assertion that the
superposition of the contributions from the large number of elementary 
collisions makes the ln$F$-ln$M$ in higher-dimensional phase space 
of relativistic heavy ion collision bending strongly upwards.

A direct implication of this finding is that, when the temperature and/or 
density are high enough so that the produced particles from individual 
elementary processes are melted into a unique system, the superposition effect
will disappear and the factorial moments will not be strongly upward-bending
any more. So the disappearance of strongly upward-bending of 
higher-dimensional factorial moments
in relativistic heavy ion collision may be taken as a signal for the formation 
of a unique system, or QGP, in this collision.

Finally, let us notice that the proposed signal for QGP formation 
has the distinguish feature that it is based on
the minimum-bias data of relativistic heavy-ion collisions. 
This has some advantage, especially for single event analysis in the future
super-high energy heavy-ion collider experiments, such as those at 
RHIC and LHC.

\noindent
{\bf Acknowledgement}

This work is supported in part by the NNSF of China
The authors are grateful to 
Evert Stenlund, Wu Yuanfang and Yang Cunbing for helpful discussions.

\bigskip \bigskip

\noindent{\Large Figure Captions}

\medskip

\noindent Fig. 1 \ Schematic plot of two-dimensional phase space partition, 

\ \ ($a)$\ self-similar; 

\ \ ($b$)\ self-affine.

\medskip

\noindent Fig. 2 \ Schematic plot of the effective partition of longitudiunal 
phase space, 

\ \ ($a$)\ in an elementary collision; 

\ \ ($b$)\ in the superposition of elementary collisions.

\medskip

\noindent Fig. 3 \ Results of model calculation with different number $A$
of elementary collisions.

\end{document}